\documentclass[conference]{IEEEtran}
\usepackage{cite}
\usepackage{amsmath,amssymb,amsfonts}
\usepackage{algorithmic}
\usepackage{graphicx}
\usepackage{textcomp}
\usepackage{xcolor}
\def\BibTeX{{\rm B\kern-.05em{\sc i\kern-.025em b}\kern-.08em
    T\kern-.1667em\lower.7ex\hbox{E}\kern-.125emX}}

\usepackage{orcidlink}
\usepackage{url}

\usepackage{breakurl}
\usepackage{hyperref}
\hypersetup{hidelinks,unicode,breaklinks}

\usepackage{calc} 
\usepackage{verbatim} 
\usepackage{balance} 

\usepackage{caption}
\usepackage{booktabs}
\usepackage{threeparttable}
\usepackage{multirow} 

\usepackage{enumitem}

\makeatletter 
\renewcommand{\@biblabel}[1]{[#1]\hfill}
\makeatother

\usepackage{tikz}
\usepackage{pgfplots}
\usetikzlibrary{calc,decorations.text,shapes,shapes.arrows,arrows,arrows.meta,bending, backgrounds}
\usepackage{ifthen}

\usepackage{ifpdf}
\ifpdf
\pdfoutput=1 
\pdftrue
\fi

\makeatletter
\def\ps@IEEEtitlepagestyle{
	\def\@oddfoot{\mycopyrightnotice}
	\def\@evenfoot{}
}
\def\mycopyrightnotice{
	{\footnotesize
		\begin{minipage}{0.8\textwidth}
			\centering
			Please cite as: Simon Liebl, Leah Lathrop, Ulrich Raithel, Andreas Aßmuth, Ian Ferguson, and Matthias Söllner, ``Analyzing the Attack Surface and Threats of Industrial Internet of Things Devices,'' \emph{International Journal On Advances in Security}, vol.~14, no.~1 and 2, pp.~59--70, 2021.
		\end{minipage}
	}
}
\makeatother

\tikzset{
	MyArrow/.style args={#1}{
		to path={let \p1 = ($(\tikztotarget)-(\tikztostart)$),
			\n1 = {int(mod(scalar(atan2(\y1,\x1))+360, 360))}, 
			\n2 = {veclen(\x1,\y1)} in \pgfextra{\typeout{\n1,\n2,\x1,\y1}} (\tikztotarget)
			node[
			#1 arrow, 
			#1 arrow head extend=1ex,
			draw,
			fill=gray!70,
			minimum height=\n2-\pgflinewidth-1,
			inner sep=0.7ex,
			rotate=\n1,%
			anchor=east,%
			]{}
	}},
	MyArrow/.default=single
}

\definecolor{blue1}{RGB}{30, 189, 247} 

\begin{document}

\title{\textbf{\LARGE Analyzing the Attack Surface and Threats of\\[-3mm]Industrial Internet of Things Devices}}

\author{\IEEEauthorblockN{\large Simon Liebl\IEEEauthorrefmark{1}\IEEEauthorrefmark{2}\orcidlink{0000-0003-1311-4202}, Leah Lathrop\IEEEauthorrefmark{1}, Ulrich Raithel\IEEEauthorrefmark{3}, Andreas Aßmuth\IEEEauthorrefmark{1}\orcidlink{0009-0002-2081-2455},\\ Ian Ferguson\IEEEauthorrefmark{2}\orcidlink{0000-0001-9866-6182}, and Matthias Söllner\IEEEauthorrefmark{1}\\[5mm]}
\IEEEauthorblockA{\IEEEauthorrefmark{1}Technical University of Applied Sciences OTH Amberg-Weiden, Amberg, Germany \\
	E-mail: {\{s.liebl $|$ l.lathrop $|$ a.assmuth $|$ m.soellner\}@oth-aw.de} \\
	\IEEEauthorrefmark{2}Abertay University, Dundee, UK\\
	E-mail: {i.ferguson@abertay.ac.uk}\\
\IEEEauthorrefmark{3}SIPOS Aktorik GmbH, Altdorf, Germany\\
E-mail: {ulrich.raithel@sipos.de}
}}

\maketitle

\begin{abstract}
The growing connectivity of industrial devices as a result of the Internet of Things is increasing the risks to Industrial Control Systems. Since attacks on such devices can also cause damage to people and machines, they must be properly secured. Therefore, a threat analysis is required in order to identify weaknesses and thus mitigate the risk. In this paper, we present a systematic and holistic procedure for analyzing the attack surface and threats of Industrial Internet of Things devices. Our approach is to consider all components including hardware, software and data, assets, threats and attacks throughout the entire product life cycle.
\end{abstract}

\renewcommand\IEEEkeywordsname{Keywords}

\section{Introduction}

The Internet of Things (IoT) is increasingly making its way into our daily lives. Smart devices are omnipresent, in smart homes, medical and infrastructure applications, and building automation. Industrial applications, for example, in manufacturing, the automotive and oil and gas industry, are other major fields of application, summarized as the Industrial Internet of Things (IIoT). The hype around the IoT and IIoT led, for example, to dozens of different platforms and, consequently, to compatibility problems. The race for the shortest time to market also led to security and privacy issues, as these topics have been neglected or even omitted entirely so far. To address the latter issues, our work in \cite{MyTA} and this extension aim to support IIoT device manufacturers and operators in identifying threats against their devices.

According to \cite{Kaspersky2020H2}, about every third Industrial Control System (ICS) computer was attacked within the second half of 2020. The situation was also exacerbated by the COVID-19 pandemic, as increasing Remote Desktop Protocol (RDP) connections also led to a rise in brute force attacks on them. A rise in network-capable Operational Technology (OT) components has been observable for years anyway \cite{Fortinet}. The main threat arises from ransomware, i.e., malware that encrypts files and demands ransom, and coinminers, i.e., malware used to mine cryptocurrencies \cite{TrendICS}. The attack on Colonial Pipeline \cite{Pipeline} showed once again that larger parts of the population can also be affected by such attacks. In addition, the attack on a U.S. water treatment facility in early 2021 \cite{Water} highlighted that ICSs in critical infrastructures are particularly at risk of targeted attacks.

As a consequence of the increasing threats, IIoT manufacturers must secure their devices to prevent such incidents. However, implementing best practices around default passwords is not enough for properly secured devices. Manufacturers, and also operators, must therefore be fully aware of threats and the resulting risks. The purpose of this paper is to support them in their threat analysis. Our goal is a holistic view of the attack surface of IIoT devices. To achieve this, all components including hardware, software, and data, assets, as well as threats and attacks should be considered throughout the entire life cycle of the device. Our research methodology can be described as follows: the first two steps are analyses of the components and assets of an IIoT device, followed by a threat and an attack categorization. By considering all assets in the threat categorization, a complete list should be enabled. Similarly, the component analysis should provide a complete attack and weakness categorization.

The remainder of this paper is structured as follows: in Section~\ref{sec:IIoT}, the different terms around the IIoT are clarified. Section~\ref{sec:RelatedWork} presents related work from other researchers as well as organizations. In Section~\ref{sec:Components}, the components of an IIoT device are decomposed into hardware, software, and data, followed by an asset analysis in Section~\ref{sec:Assets}. In Section \ref{sec:Threats}, a threat categorization is presented that consists of nearly 50 categories organized into 10 groups. A similar categorization of attacks and weaknesses throughout the life cycle of an IIoT device is introduced in Section \ref{sec:Attacks}. Section \ref{sec:Procedure} recommends six steps for a systematic threat analysis of IIoT devices. The paper ends with conclusions in Section \ref{sec:Conclusions}.

\section{The Industrial Internet of Things}
\label{sec:IIoT}

\begin{figure}[h!]
	\centering
	\resizebox{0.43\textwidth}{!}{
		\begin{tikzpicture}	
			\shade[left color=gray!10,right color=gray!30] (-1.7,-1.7) rectangle +(5.15,3.4);
			\fill[blue1, fill opacity=0.4] (0,0) circle (1.5cm);
			\fill[yellow, fill opacity=0.4] (1.75,0) circle (1.5cm);
			\fill[blue1!90, fill opacity=0.7, draw=black, draw opacity=0.5] (0.5,0) circle (0.85cm);
			
			\node[text width=0.5cm, align=center] at (-1.4, 1.45) {\small IT};
			\node[text width=0.5cm, align=center] at (3.15, 1.45) {\small OT};
			
			\node[text width=0.5cm, align=center] at (-1.1, 0) {\small IoT};
			\node[text width=0.5cm, align=center] at (0.5, 0) {\small IIoT};
			\node[text width=0.5cm, align=center] at (2.25, 0) {\small CPS};
		\end{tikzpicture}
	}
	\caption{Relationship of the various terms, adopted from \cite{IIoTChallOpDir}.}
	\label{fig:Terms}
\end{figure}
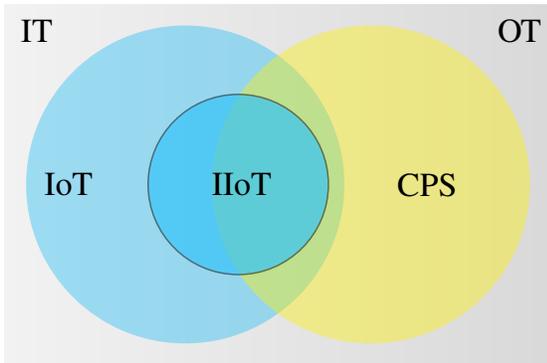

This section briefly explains the terms IoT, IIoT, Cyber-Physical System (CPS), ICS, Information Technology (IT), and OT and how they are related (see Figure~\ref{fig:Terms}).

The IoT is a network of connected devices, which are sensors and/or actuators fulfilling a specific function. The infrastructure enables communication with other equipment along with the storage, processing, and distribution of data to other systems and users \cite{IoTDefnTax}. Cloud services realize universal accessible utility services, e.g., for device management, and centralized data processing for analytics, which may be supported by Artificial Intelligence (AI). The gained knowledge can be made available to users, other systems and the devices themselves. Use cases span many domains, such as consumer applications (e.g., smart home), commercial (e.g., medical and healthcare, transportation), and infrastructure applications (e.g., smart grid).

The IIoT is a part of the IoT, but differentiates in some aspects. The IIoT integrates the previously separated areas of IT and OT by connecting OT components, such as machines and control systems, with IT systems and business processes \cite{IIoTChallOpDir}. The integration is accomplished, for instance, through edge devices and gateways that enable processing in the cloud. The leading use cases of the IIoT are operational intelligence, asset monitoring, and predictive maintenance \cite{IIoTUseCases}. The goals are, among others, to increase productivity, improve safety, gain flexibility and agility, and reduce energy consumption. It should be noted that use cases, services, and communication in the IIoT are machine-oriented, while these are human-centered in many IoT applications, such as in consumer IoT \cite{IIoTChallOpDir}. 

The OT components listed above are usually installed within an ICS, which is structured into several layers. In the Purdue reference model, level 0 describes the physical process, which is sensed and manipulated by sensors and actuators and controlled by, for example, Programmable Logic Controllers (PLCs) in level 1. Level 2 includes Human-Machine Interfaces (HMIs) and the Supervisory Control And Data Acquisition (SCADA) system to provide operators with aggregated information. Layers above contain backup servers and Enterprise Resource Planning (ERP) systems, among others.

The last term that needs to be clarified is CPS. These can be found in the IoT/IIoT and in an ICS. Their primary task is the control of a physical process in the real world using sensors and actuators. Furthermore, they are equipped with network capability. Another characteristic is that they require real-time interaction with the physical world \cite{IIoTAnalysisFW}.

Different objectives and requirements emerge from the characteristics of CPSs and ICSs. Besides integrity and availability, the security goals for IoT devices are centered on confidentiality and privacy, as personal data, such as health data, is processed. IIoT devices, in contrast, focus additionally on safety and the impact on the environment and society \cite{OTinICS}. In industrial plants, humans work with heavy machinery in a confined workspace. An accident can potentially cause injury, death, damaged production equipment or environmental disasters. The impact of an IIoT failure may be worse in critical infrastructures, such as energy and water supply, food, and health, as large parts of the population may be affected.

To sum up, the IIoT allows the processing of data, e.g., produced by CPSs in an ICS, in the cloud. This is realized by connected edge devices, gateways, and OT components. The control of physical processes leads to further requirements such as safety. The increased connectivity and the high requirements result in additional threats and increased risk to IIoT devices, which is further discussed in Section \ref{sec:Attacks}.

\section{Related Work}
\label{sec:RelatedWork}

The need for action in the area of IoT security was recognized by experts a long time ago \cite{SurveyIIoTSec}. In recent years, this has also been identified by government institutions and industry. Nevertheless, manufacturers of embedded systems still struggle to integrate security features or even to work according to the principle of security by design.

Varga et al. \cite{ThreatsAutomation} discuss IoT threats in the automation domain. Based on an IoT architecture consisting of the four layers sensors and actuators, networking, data processing, and application, the authors present different threats and the required countermeasures. In \cite{SecAnalysis}, Atamli et al. describe a threat-based security analysis that focuses on the three use cases power management, smart car, and smart healthcare system. Initially, they discuss sources of threats and classify attack vectors. Afterwards, the security and privacy impact of attacks in the area of the listed use cases is described. Last, desirable security and privacy properties are defined. Abomhara et al. \cite{IoTattacks} provide background information on IoT devices and services, threats, attacks, and security and privacy goals. Subsequently, the motivation of attacks and a classification of possible intruders are presented. In \cite{CIsecAnalysis}, Wurm et al. conducted  security analyses on a consumer IoT and an IIoT device and demonstrated how these devices could be exploited.  

Dozens of organizations and government institutions, such as the Industrial Internet Consortium (IIC), the Cloud Security Alliance (CSA), and the US National Institute of Standards and Technology (NIST), have published guidelines and best practices for IoT security. Particularly noteworthy is \cite{Mapping}, which brought together about 100 documents from 50 different organizations, resulting in 13 points for recommended action. Furthermore, the contributions of the European Union Agency for Cybersecurity (ENISA) and the German Federal Office for Information Security (BSI) are recommended. The former have published several analyses and recommendations for the various areas of the IoT \cite{ENISAIoT}. For example, the baseline security recommendations for IoT \cite{ENISABaseline} provides a threat taxonomy, attack scenarios and a list of security measures. The BSI annually publishes an Information Security Management System (ISMS), the so-called IT-Grundschutz Compendium, which also considers ICS components, embedded systems, and IoT devices, among others. In addition to organizational aspects, technical issues are also addressed, including a list of threats and necessary security requirements.

There are numerous papers and guidelines describing the threats to IoT devices. However, the threats are often described only in general terms and in a jumbled manner. For example, the aforementioned threat taxonomy by ENISA lists 25 threats, but attack techniques (e.g., replay of messages), weaknesses (e.g., software vulnerabilities), and threat consequences (e.g., sensitive information leaking) are considered without further distinction. Systematic and holistic approaches are needed to enable the identification of all attack vectors of IIoT devices by their manufacturers. It is this shortcoming that we wish to address in this work.

\section{Components of an IIoT Device}
\label{sec:Components}

\begin{figure}[ht!]
	\centering
	\resizebox{0.5\textwidth}{!}{
		\begin{tikzpicture}
			\def \radius {2.5}
			\fill[color=blue1!60] (0,0) circle (1mm);
			\filldraw[color=blue1!30] (0,0) -- ({\radius*cos(0)},{\radius*sin(0)}) arc (0:90:\radius) -- cycle;
			\filldraw[color=blue1!30] (0,0) -- ({\radius*cos(330)},{\radius*sin(330)}) arc (330:360:\radius) -- cycle;
			\filldraw[color=blue1!45] (0,0) -- ({\radius*cos(90)},{\radius*sin(90)}) arc (90:210:\radius) -- cycle;
			\filldraw[color=blue1!15] (0,0) -- ({\radius*cos(210)},{\radius*sin(210)}) arc (210:330:\radius) -- cycle;
			
			\pgfmathtruncatemacro\ang{-30+12}
			\pgfmathtruncatemacro\pad{24}
			\def \hei {3mm}
			\def \heiL {3.5mm}
			\def \off {1.35}
			\fill[color=blue1!60] ({\radius*cos(\ang)},{\radius*sin(\ang)}) circle (3mm);
			\node at ({\radius*cos(\ang)},{\radius*sin(\ang)}) {\includegraphics[height=\heiL]{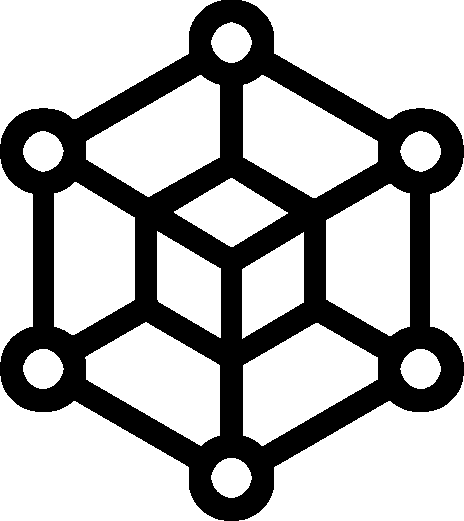}};
			\node[text width=2cm, align=left] at ({\radius*cos(\ang)+\off},{\radius*sin(\ang)}) {\small Application};
			\pgfmathtruncatemacro\ang{\ang+\pad};
			\fill[color=blue1!60] ({\radius*cos(\ang)},{\radius*sin(\ang)}) circle (3mm);
			\node at ({\radius*cos(\ang)},{\radius*sin(\ang)}) {\includegraphics[height=\heiL]{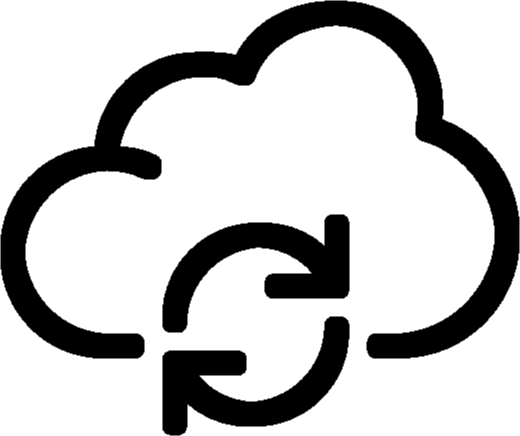}};
			\node[text width=2cm, align=left] at ({\radius*cos(\ang)+\off},{\radius*sin(\ang)}) {\small Services\,/\,API};
			\pgfmathtruncatemacro\ang{\ang+\pad};
			\fill[color=blue1!60] ({\radius*cos(\ang)},{\radius*sin(\ang)}) circle (3mm);
			\node at ({\radius*cos(\ang)},{\radius*sin(\ang)}) {\includegraphics[height=\heiL]{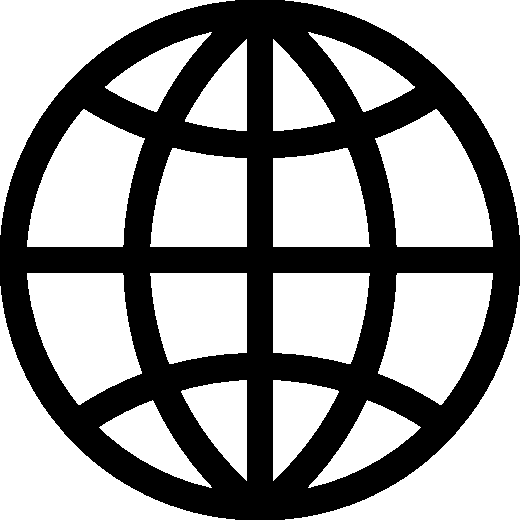}};
			\node[text width=2cm, align=left] at ({\radius*cos(\ang)+\off},{\radius*sin(\ang)}) {\small Connectivity};
			\pgfmathtruncatemacro\ang{\ang+\pad};
			\fill[color=blue1!60] ({\radius*cos(\ang)},{\radius*sin(\ang)}) circle (3mm);
			\node at ({\radius*cos(\ang)},{\radius*sin(\ang)}) {\includegraphics[height=\heiL]{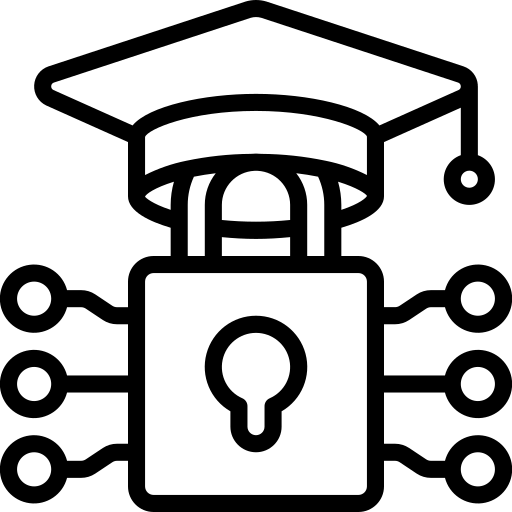}};
			\node[text width=2cm, align=left] at ({\radius*cos(\ang)+\off},{\radius*sin(\ang)}) {\small Cryptography};
			\pgfmathtruncatemacro\ang{\ang+\pad};
			\fill[color=blue1!60] ({\radius*cos(\ang)},{\radius*sin(\ang)}) circle (3mm);
			\node at ({\radius*cos(\ang)},{\radius*sin(\ang)}) {\includegraphics[height=\hei]{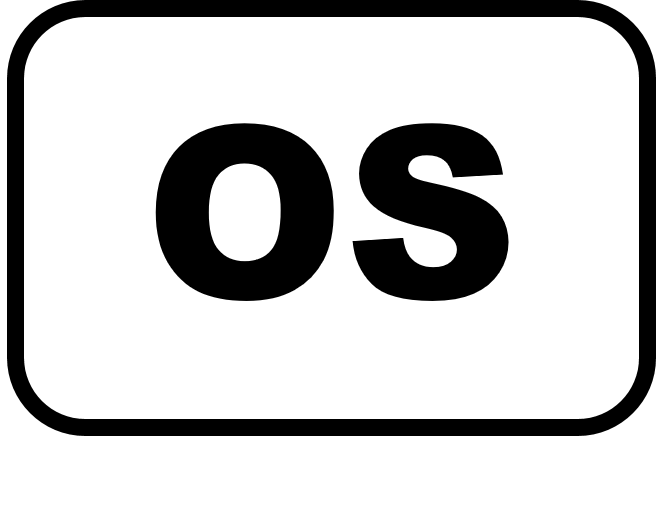}};
			\node[text width=2cm, align=left] at ({\radius*cos(\ang)+\off-0.05},{\radius*sin(\ang)+0.3}) {\small Firmware\,/\,OS/RTOS};
			\pgfmathtruncatemacro\ang{\ang+\pad};
			
			\def \off {-1.35}
			\fill[color=blue1!60] ({\radius*cos(\ang)},{\radius*sin(\ang)}) circle (3mm);
			\node at ({\radius*cos(\ang)},{\radius*sin(\ang)}) {\includegraphics[height=\hei]{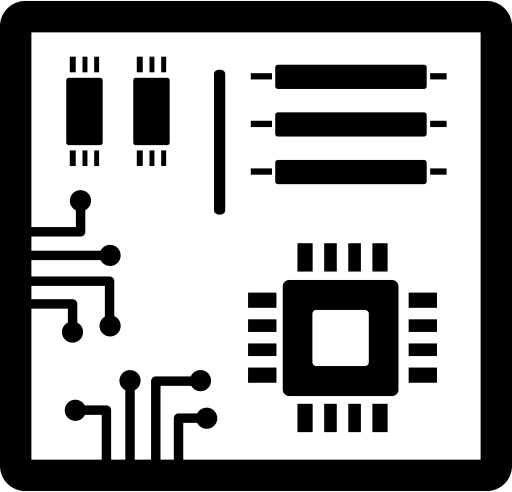}};
			\node[text width=2cm, align=right] at ({\radius*cos(\ang)+\off+0.05},{\radius*sin(\ang)+0.3}) {\small Circuit Board};
			\pgfmathtruncatemacro\ang{\ang+\pad};
			\fill[color=blue1!60] ({\radius*cos(\ang)},{\radius*sin(\ang)}) circle (3mm);
			\node at ({\radius*cos(\ang)},{\radius*sin(\ang)}) {\includegraphics[height=\heiL]{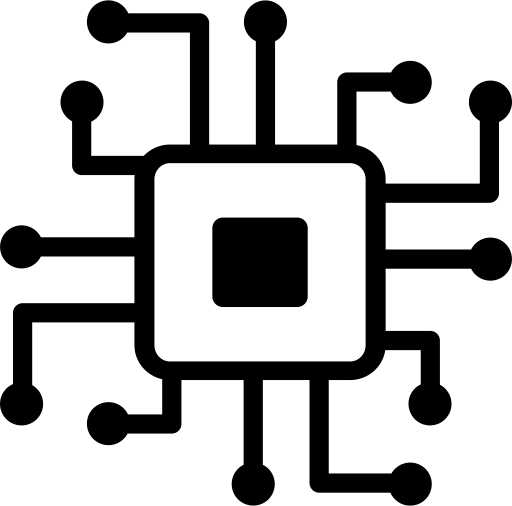}};
			\node[text width=2cm, align=right] at ({\radius*cos(\ang)+\off},{\radius*sin(\ang)}) {\small Microprocessor};
			\pgfmathtruncatemacro\ang{\ang+\pad};
			\fill[color=blue1!60] ({\radius*cos(\ang)},{\radius*sin(\ang)}) circle (3mm);
			\node at ({\radius*cos(\ang)},{\radius*sin(\ang)}) {\includegraphics[height=\hei]{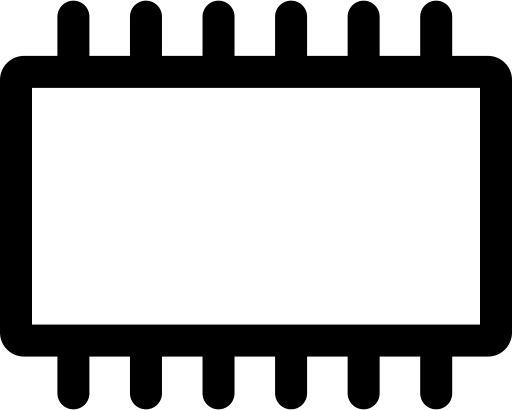}};
			\node[text width=2cm, align=right] at ({\radius*cos(\ang)+\off},{\radius*sin(\ang)}) {\small Memory};
			\pgfmathtruncatemacro\ang{\ang+\pad};
			\fill[color=blue1!60] ({\radius*cos(\ang)},{\radius*sin(\ang)}) circle (3mm);
			\node at ({\radius*cos(\ang)},{\radius*sin(\ang)}) {\includegraphics[height=\hei]{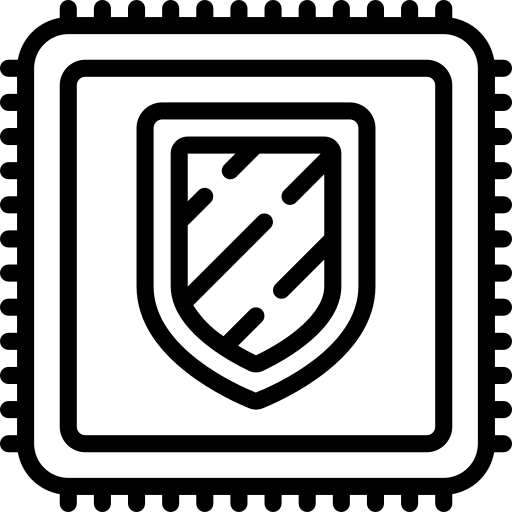}};
			\node[text width=2cm, align=right] at ({\radius*cos(\ang)+\off},{\radius*sin(\ang)}) {\small Security Chip};
			\pgfmathtruncatemacro\ang{\ang+\pad};
			\fill[color=blue1!60] ({\radius*cos(\ang)},{\radius*sin(\ang)}) circle (3mm);
			\node at ({\radius*cos(\ang)},{\radius*sin(\ang)}) {\includegraphics[height=\heiL]{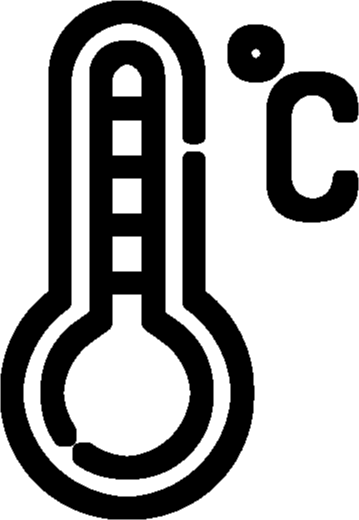}};
			\node[text width=2cm, align=right] at ({\radius*cos(\ang)+\off-0.15},{\radius*sin(\ang)}) {\small Sensor\,/\,Actuator};
			\pgfmathtruncatemacro\ang{\ang+\pad};
			
			\def \off {-1.6}
			\fill[color=blue1!60] ({\radius*cos(\ang)},{\radius*sin(\ang)}) circle (3mm);
			\node at ({\radius*cos(\ang)},{\radius*sin(\ang)}) {\includegraphics[height=\heiL]{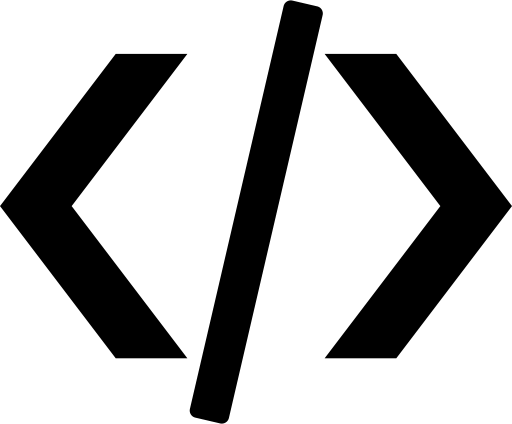}};
			\node[text width=2.5cm, align=right] at ({\radius*cos(\ang)+\off},{\radius*sin(\ang)}) {\small Code};
			\pgfmathtruncatemacro\ang{\ang+\pad};
			\fill[color=blue1!60] ({\radius*cos(\ang)},{\radius*sin(\ang)}) circle (3mm);
			\node at ({\radius*cos(\ang)},{\radius*sin(\ang)}) {\includegraphics[height=\heiL]{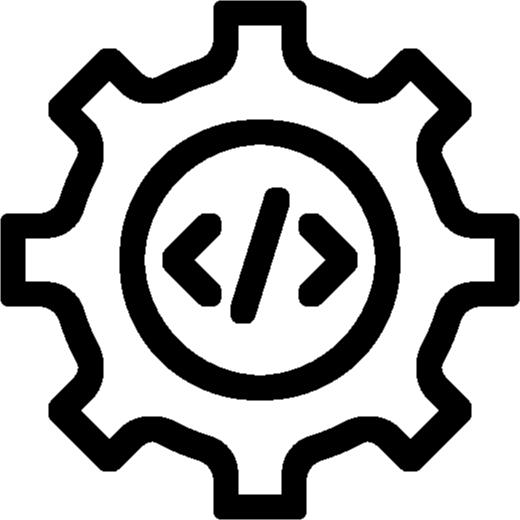}};
			\node[text width=2.5cm, align=right] at ({\radius*cos(\ang)+\off},{\radius*sin(\ang)-0.1}) {\small Configuration Data};
			\pgfmathtruncatemacro\ang{\ang+\pad};
			\fill[color=blue1!60] ({\radius*cos(\ang)},{\radius*sin(\ang)}) circle (3mm);
			\node at ({\radius*cos(\ang)},{\radius*sin(\ang)}) {\includegraphics[height=\heiL]{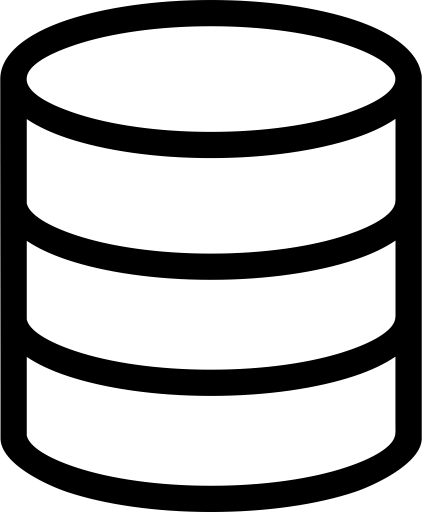}};
			\node[text width=2.5cm, align=right] at ({\radius*cos(\ang)+\off+1.45},{\radius*sin(\ang)-0.5}) {\small Application Data};
			\pgfmathtruncatemacro\ang{\ang+\pad};
			\def \off {1.6}
			\fill[color=blue1!60] ({\radius*cos(\ang)},{\radius*sin(\ang)}) circle (3mm);
			\node at ({\radius*cos(\ang)},{\radius*sin(\ang)}) {\includegraphics[height=\heiL]{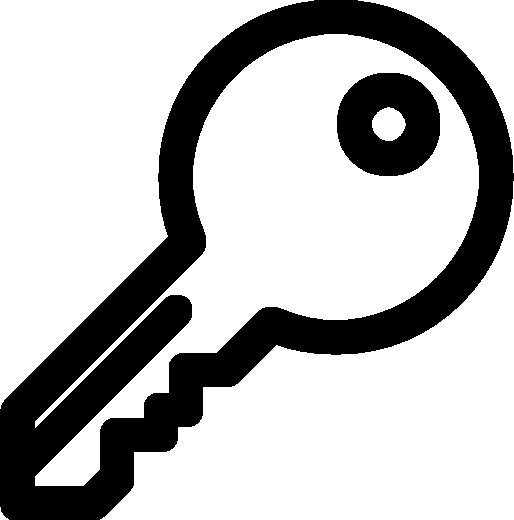}};
			\node[text width=3.9cm, align=left] at ({\radius*cos(\ang)+\off+0.7},{\radius*sin(\ang)-0.1}) {\small Access Data\,/\,Keys};
			\pgfmathtruncatemacro\ang{\ang+\pad};
			\fill[color=blue1!60] ({\radius*cos(\ang)},{\radius*sin(\ang)}) circle (3mm);
			\node at ({\radius*cos(\ang)},{\radius*sin(\ang)}) {\includegraphics[height=\heiL]{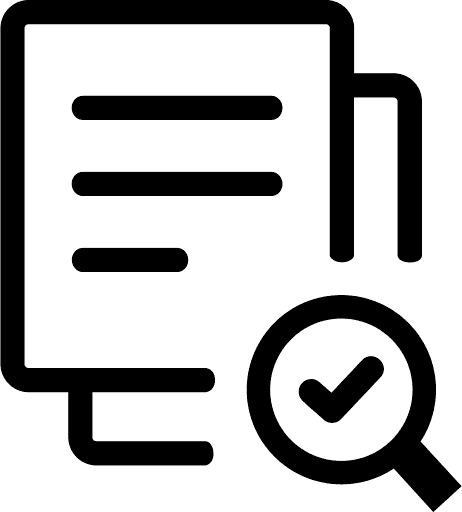}};
			\node[text width=2.5cm, align=left] at ({\radius*cos(\ang)+\off},{\radius*sin(\ang)}) {\small Log Data};
			
			\def \inrad {0.75}
			\def \inhei {5mm}
			\node at ({\inrad*cos(30)},{\inrad*sin(30)}) {\includegraphics[height=\inhei]{code.png}};
			\node at ({\inrad*cos(150)},{\inrad*sin(150)}) {\includegraphics[height=\inhei]{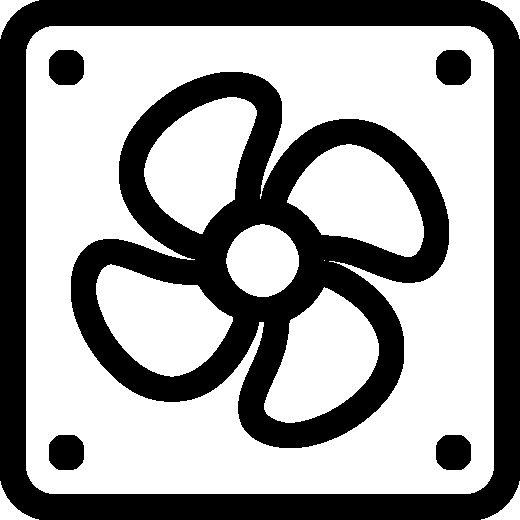}};
			\node at ({\inrad*cos(270)},{\inrad*sin(270)}) {\includegraphics[height=\inhei]{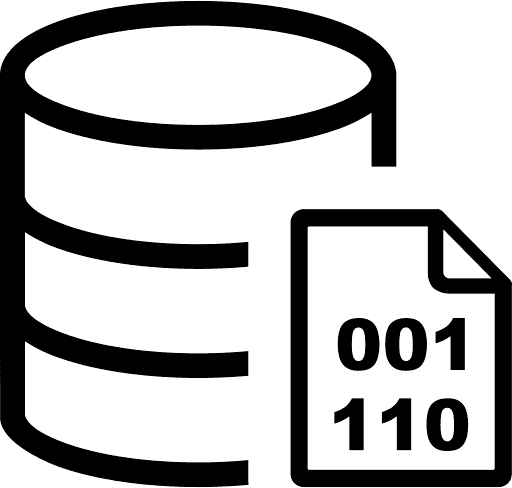}};
			
			\def \scal {0.75}
			\def \sx {0}
			\def \sy {1.75}
			\draw [draw=blue1!45,postaction={decorate,decoration={text along path,text align=center,reverse path,text={Hardware}}}] (\sx,\sy) to [bend right=45] ({\sx*cos(120)-\sy*sin(120)},{\sx*sin(120)+\sy*cos(120)});
			
			\draw [draw=blue1!30,postaction={decorate,decoration={text along path,text align=center,reverse path,text={Software}}}] ({\sx*cos(-120)-\sy*sin(-120)},{\sx*sin(-120)+\sy*cos(-120)}) to [bend right=45] (\sx,\sy);
			
			\def \sy {1.95}
			\draw [draw=blue1!15,postaction={decorate,decoration={text along path,text align=center,text={Data}}}] ({\sx*cos(120)-\sy*sin(120)},{\sx*sin(120)+\sy*cos(120)}) to [bend right=45] ({\sx*cos(-120)-\sy*sin(-120)},{\sx*sin(-120)+\sy*cos(-120)});
			
	\end{tikzpicture}}
	\caption{Hardware and software components and types of data that IIoT devices may contain.}
	\label{fig:DeviceComp}
\end{figure}

To enable the full analysis of attack surfaces, a breakdown of the components of a typical device is presented in the following section. The components can be grouped into the three categories hardware, software, and data, which are described in the following subsections in detail. Figure~\ref{fig:DeviceComp} presents an overview of the components an IIoT device may contain.

\subsection{Hardware}

The first component that comes to mind is the enclosure. It must be suitable for the environment and may have to be explosion-, water-, and dust-proof. Appearance, size, and usability are particularly important in the consumer sector, but the industrial field also appreciates these properties. Plant operators, for example, prefer simple and space-saving installation in the switching cabinet and quick familiarization with handling by employees. In critical applications, tamper protection is used to detect modifications to the device. 

A central component of the interior are Printed Circuit Boards (PCBs). They are often the bridge between the various mechanical and electrical parts of a device and also connect the countless electrical components on a PCB such as controllers, Integrated Circuits (ICs), oscillators, fuses, and basic electrical elements.

The core components on a PCB are processors. Several types are available, each with its own benefits and drawbacks. Among others, there are microprocessors, microcontrollers, Application-Specific Integrated Circuits (ASICs), and Field-Programmable Gate Arrays (FPGAs). Microprocessors ship in a single IC, which vice versa may contain multiple microprocessors in case of a multi-core design. They can be made for general-purpose or specialized on a specific task, e.g., signal, graphics, and physics processing. In addition to the microprocessor, dozens of peripherals are integrated into microcontroller ICs, such as memory, analog and digital inputs and outputs, serial communication interfaces, a Real-Time Clock (RTC) and in-circuit debug support. Increasingly, security features such as a Trusted Execution Environment (TEE), a True Random Number Generator (TRNG), a cryptography accelerator, and a Physical Unclonable Function (PUF) are also being embedded. ASICs are customized for a certain task and their advantages include, for instance, greater performance and optimized size. Unlike ASICs, FPGAs can be updated after production and are more cost-effective, especially for smaller quantities. IoT devices usually employ microcontrollers as their main Central Processing Unit (CPU) because they are feature-rich and are still low-priced and compact. The average CPU has a single core and the clock speed is in the double-digit MHz range, which drastically limits the performance compared to desktop CPUs in IT systems.

Memory can be integrated into the microcontroller, placed as separate IC on the PCB or connected by slots in the enclosure. The main memory is typically Static Random Access Memory (SRAM) or Dynamic Random Access Memory (DRAM). There are different technologies for data storage, for example, FLASH, Electrically Erasable Programmable Read-Only Memory (EEPROM), and One-Time Programmable (OTP) memory. Some security-focused microcontrollers include a on-the-fly encryption engine that enables secure storage on external ICs. Removable storage technologies, such as SD cards or USB sticks, are used, for example, to export user data or import user applications. The total main memory is usually in the kilobyte range, the data storage reaches the megabyte range.

Like memory, a security chip can also be integrated into a microcontroller or placed on the PCB. Security chip is used as umbrella term for secure cryptoprocessors, which can be a Trusted Platform Module (TPM) or a Secure Element (SE). An exception are Hardware Security Modules (HSMs) that are integrated via an external module. Their capabilities differ in certain functions, although the common basic idea is to outsource all cryptographic operations to a tamper-proof co-processor. 

Another substantial part of IIoT devices are input and output components including sensors and actuators. The power supply is also a type of input component, as it supplies the device with power via a cable connection, battery, or solar panel, among others. User input interfaces may include simple switches or utilize more advanced human input devices such as keyboards or touchpads. According to Sikder et al.\cite{SurveySensorThreats}, the various sensor can be categorized into environmental sensors (e.g., audio, image, temperature and humidity sensor), position sensors (e.g., inductive, ultrasonic, proximity, and magnetic sensor) and motion sensors (e.g., flow sensor, gyroscope, accelerometer); we extended this list with industrial sensors, summarized as process sensors (e.g., current, pressure, and chemical sensor). The output components can be grouped in visual outputs (e.g., LEDs, display), audio outputs (e.g., loudspeaker), power outputs (e.g., relay, power electronics), and actuators (e.g., electric and pneumatic actuator).

The last group of components in this list are connectivity elements. This includes conductive paths on the PCB and wires within the device. The numerous communication interfaces of a system may require controllers, connectors, or antennas.

\subsection{Software}

The firmware is the linking component between hardware and software of an embedded system, as it provides software with low-level access to the hardware. Simple embedded devices often have no underlying Operating System (OS); therefore, they run only the firmware, which is known as bare metal program. The most utilized OS in IoT devices is Linux \cite{OSes}, followed by FreeRTOS, an open source Real-Time OS (RTOS). In addition to real-time capability, some CPSs in safety-critical applications may require the fulfillment of further standards such as IEC 61508. Specialized OSs have been developed to comply with these requirements, for example, SAFERTOS achieved the highest Safety Integrity Level (SIL) of IEC 61508 for a software-only component, i.e., SIL 3. 
 
Although hardware-supported cryptography in the form of cryptographic accelerators or security chips is more efficient than software libraries, most IoT devices use mainly software-based cryptography. The main reasons for this are that crypto-hardware is hardly available and software libraries often do not exploit available hardware for portability reasons \cite{CryptoHW}. There is a wide range of fee-based as well as cost-free crypto libraries available. However, there are some challenges that complicate their use. First, only a few of them can easily be adapted to systems without an OS, which is about every fourth IoT device \cite{OSes}. Second, the required storage space exceeds the frequently limited memory. Finally, the execution of cryptographic algorithms on low-power devices is highly time-consuming and, therefore, compromises other requirements such as real-time capability and usability. These challenges are addressed in the research field lightweight cryptography in order to provide efficient and storage-saving cryptographic software libraries on all devices.

Connectivity is one of the major topics addressed during the design of an IoT device as it defines how a device interacts with other systems as well as users. A wide range of communication protocol stacks are used for the various applications. There are stacks for local communication (e.g., USB), Internet communication (e.g., TCP/IP), and automation processes (e.g., Modbus). Currently, there is a trend towards Ethernet-based automation protocols to take advantage of synergy. IIoT devices usually implement several protocol stacks for compatibility reasons. This results in devices supporting legacy protocols such as PROFIBUS and HART as well as more recent ones such as PROFINET, OPC UA, MQTT, and LoRa.

Smart services unlock the value of the IoT. Countless devices can thus connect with each other and create an elaborated decision. At the same time, they can be centrally monitored and controlled. One central service is device management, which includes device registration, organization, inspection, and software and firmware updates. This may be accompanied by several other services, for instance, monitoring, logging, and attestation services. Frequently, a local service is also required for configuration and control that is usually only available in the local network. Therefore, many devices implement an embedded web server or a mobile app Application Programming Interface (API) for this purpose. 

Last, every device runs its own specific application. Depending on the use case, this might be providing sensor values, controlling an actuator, or bundling messages from multiple smart sensors. Edge devices may also perform data pre-processing or minor analytics. Additionally, some devices allow users to execute their own programs and code. 

\subsection{Data}

IIoT devices hold various types of data. First of all, any code including the firmware and application can be considered as data. Code is preferably stored in a FLASH memory or, if it is not available, in an EEPROM. 

During device setup, many options need to be configured. Configuration data can include network settings, environment and calibration parameter, sensor and actuator settings, and machine learning parameters. It is stored in an EEPROM and optionally also stored in the cloud as backup.

Application data is specific to the device. This can be collected input data, such as sensor values, and produced output data, such as analysis results. 

Devices that implement basic security controls require authentication data and cryptographic keys. Local access to the device via display, browser, or mobile app should only be allowed after entering valid credentials. The user database is often stored locally and contains the names of the respective users with passwords or the derived hash values. Cryptographic keys are, for example, required to securely communicate with any other entity such as a cloud service. Not only secret keys are stored, but also public keys, for instance the manufacturer's public key, to be able to verify firmware signatures. User databases need to be updatable and are therefore stored in an EEPROM, whereas public keys may be required to be tamper-proof and consequently stored, for example, in an OTP memory.

IoT devices log relevant events for system diagnosis. These include application-specific events as well as security-relevant events such as login attempts, for example. Unfortunately, in practice, some devices also log highly sensitive data such as passwords.

\section{Assets of an IIoT Device}
\label{sec:Assets}

This section emphasizes the assets of an IIoT device based on the previously gathered components. We identified three major stakeholders in a device: manufacturers, owners, and users. By putting oneself in the shoes of the particular group, the interests and assets worth protecting can be identified. This, in turn, contributes to the understanding of what attackers might be targeting. Figure~\ref{fig:Assets} summarizes the assets for each stakeholder as well as common ones highlighted in the arrow.

\renewcommand*\labelitemii{\textbullet}
\begin{figure}[t!]
	\centering
	\resizebox{0.5\textwidth}{!}{
	\begin{tikzpicture}	
		\fill[blue1!30] (-1.85,-1.95) rectangle (2.15,6.85);
		\fill[blue1!45] (0.15,5.25) circle (1.4cm);
		\node at (0.15,5.75) {\includegraphics[height=13mm]{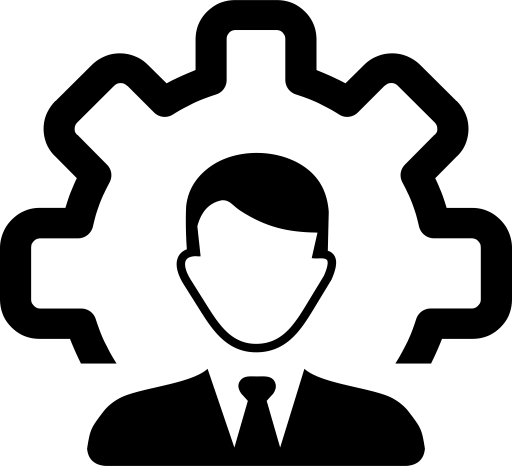}};
		\node[text width=3cm, align=center] at (0.15, 4.8) {\large Manufacturer};
		\renewcommand{\labelitemii}{\scriptsize$\blacktriangleright$}
		\node[text width=4cm, align=left, text depth=3.5cm] at (0-0.05,2) {
			\begin{itemize}
				\item Intellectual Property
				\begin{itemize}[labelindent=3mm,labelsep=1.5mm]
					\item Hardware
					\item Software
					\item Process
				\end{itemize}
				\item Liability \& Reputation
			\end{itemize}
		};
	
		\fill[blue1!30] (2.3,-1.95) rectangle (6.3,6.85);
		\fill[blue1!45] (4.3,5.25) circle (1.4cm);
		\node at (4.3,5.75) {\includegraphics[height=13mm]{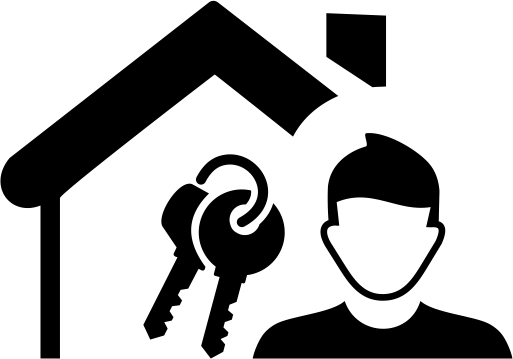}};
		\node[text width=3cm, align=center] at (4.3, 4.8) {\large Owner};
		\node[text width=4.05cm, align=left, text depth=3.5cm] at (4.15-0.05,2) {
			\begin{itemize}
				\item Physical Property
				\begin{itemize}[labelindent=0mm,labelsep=1.5mm]
					\setlength{\itemindent}{-0.5mm}
					\item Device\,/\,Environment
				\end{itemize}
				\item Virtual Property
				\begin{itemize}[labelindent=0mm,labelsep=1.5mm]
					\setlength{\itemindent}{-0.5mm}
					\item Generated Data
					\item Code\,/\,Configuration
				\end{itemize}
			\end{itemize}
		};
	
		\fill[blue1!30] (6.45,-1.95) rectangle (10.45,6.85);
		\fill[blue1!45] (8.45,5.25) circle (1.4cm);
		\node at (8.45,5.75) {\includegraphics[height=13mm]{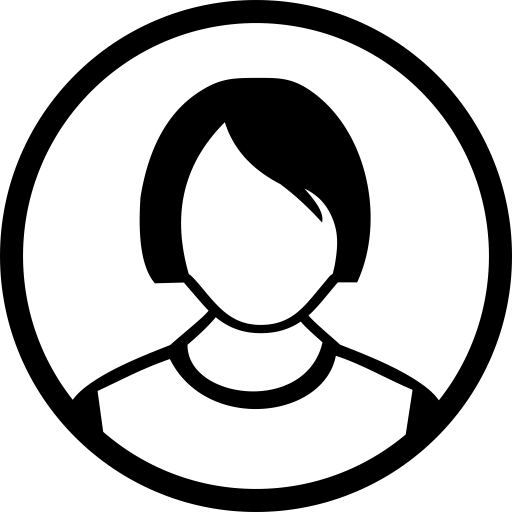}};
		\node[text width=3cm, align=center] at (8.45, 4.8) {\large User};
		\node[text width=4cm, align=left, text depth=3.5cm] at (8.3-0.05,2) {
			\begin{itemize}
				\item Health
				\item Information\,\&\,Privacy
				\begin{itemize}[labelindent=0mm,labelsep=1.5mm]
					\setlength{\itemindent}{-0.5mm}
					\item Behavioral Data
					\item Health Data
					\item Audio\,/\,Video Data
				\end{itemize}
			\end{itemize}
		};
	
		\fill[blue1!45] (-1.5,-0.25) -- (0,-1.75) -- (0, 1.25) -- cycle;
		\fill[blue1!45] (0,-1.25) rectangle (8.6,0.75);
		\fill[blue1!45] (10.1,-0.25) -- (8.6,-1.75) -- (8.6, 1.25) -- cycle;
		\node[text width=4cm, align=left] at (2.15,-0.15) {
			\begin{itemize}
				\item Functionality\,\&\,Safety
				\begin{itemize}[labelindent=0mm,labelsep=1.5mm]
					\setlength{\itemindent}{-0.5mm}
					\item Logical Operations
					\item Physical Operations
				\end{itemize}
			\end{itemize}
		};
		\node[text width=4cm, align=left] at (6.15,-0.15) {
			\begin{itemize}
				\item Authentication Data
				\begin{itemize}[labelindent=0mm,labelsep=1.5mm]
					\setlength{\itemindent}{-0.5mm}
					\item Passwords
					\item Cryptographic Keys
				\end{itemize}
			\end{itemize}
		};
		\node[text width=1cm, align=center] at (10.4, 4.8) {\phantom{t}};
	\end{tikzpicture}
	}
	\caption{Assets from the perspective of manufacturers, owners, and users as well as their common objectives.}
	\label{fig:Assets}
\end{figure}
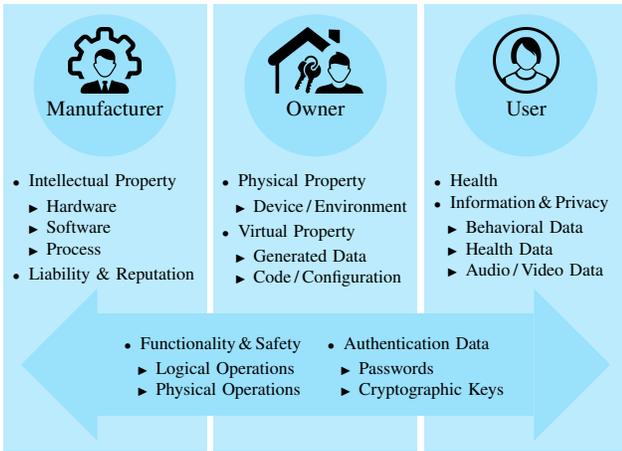

First, the assets of each group starting with manufacturers are considered and then the common ones. A primary asset of manufacturers is their Intellectual Property (IP), which can be further distinguished into the domains hardware, software, and process. Hardware IP includes, for instance, the design of an ASIC or PCB and mechanical components. Manufacturers put a lot of effort in optimizing the hardware design for energy usage, heat flow, and size. Consequently, hardware is one of the assets worth protecting in order to have an advantage over competitors. The same applies to software IP. Developing the firmware, inventing application-specific algorithms, or designing machine learning models is time-consuming. Therefore, manufacturers want to protect their software against copying, theft, or publication. Process IP means the overall concept or a unique solution for a specific problem that might be worth protecting. Two further assets are liability and reputation. Manufacturers have a certain responsibility to ensure that no unexpected incidents, such as physical injury or property damage, occur. Any incidents could also have an impact on their company's image.

Next, the assets are considered from the owner's perspective, the person or organization that purchased the device. First of all, owners want to protect their physical property, which is the newly acquired device and also all other belongings. Second, the owner has virtual property. This includes, for instance, code for custom applications and (configuration) data stored on the device. Additionally, devices produce data during operation. This includes collected sensor values and also meta information such as the availability of a device. Furthermore, owners want to prevent or detect tampering with the device by users.

Users foremost interest when using the device is their health. It must neither be endangered by one-time events nor by long-term use of the device. Users also want to protect their data on the device, which they actively stored on it or which was generated during the use of it. Especially IoT devices with dozens of sensors might collect audio and video, health, and behavioral data that could compromise the privacy of users. 

Last, the assets and objectives that all stakeholders have in common. Functionality and safety are important attributes of an IoT device. This includes logical operations, such as successful firmware updates or changing of settings, and physical operations, such as moving an actuator or cutting off electricity if a human is present. If these requirements are not achieved, this can have consequences for the manufacturer's reputation, destroy production facilities and cause downtime, or injure users. 

\section{Threat Categorization}
\label{sec:Threats}

\begin{figure*}[h!]
	\centering
	\resizebox{\textwidth}{!}{
		\includegraphics{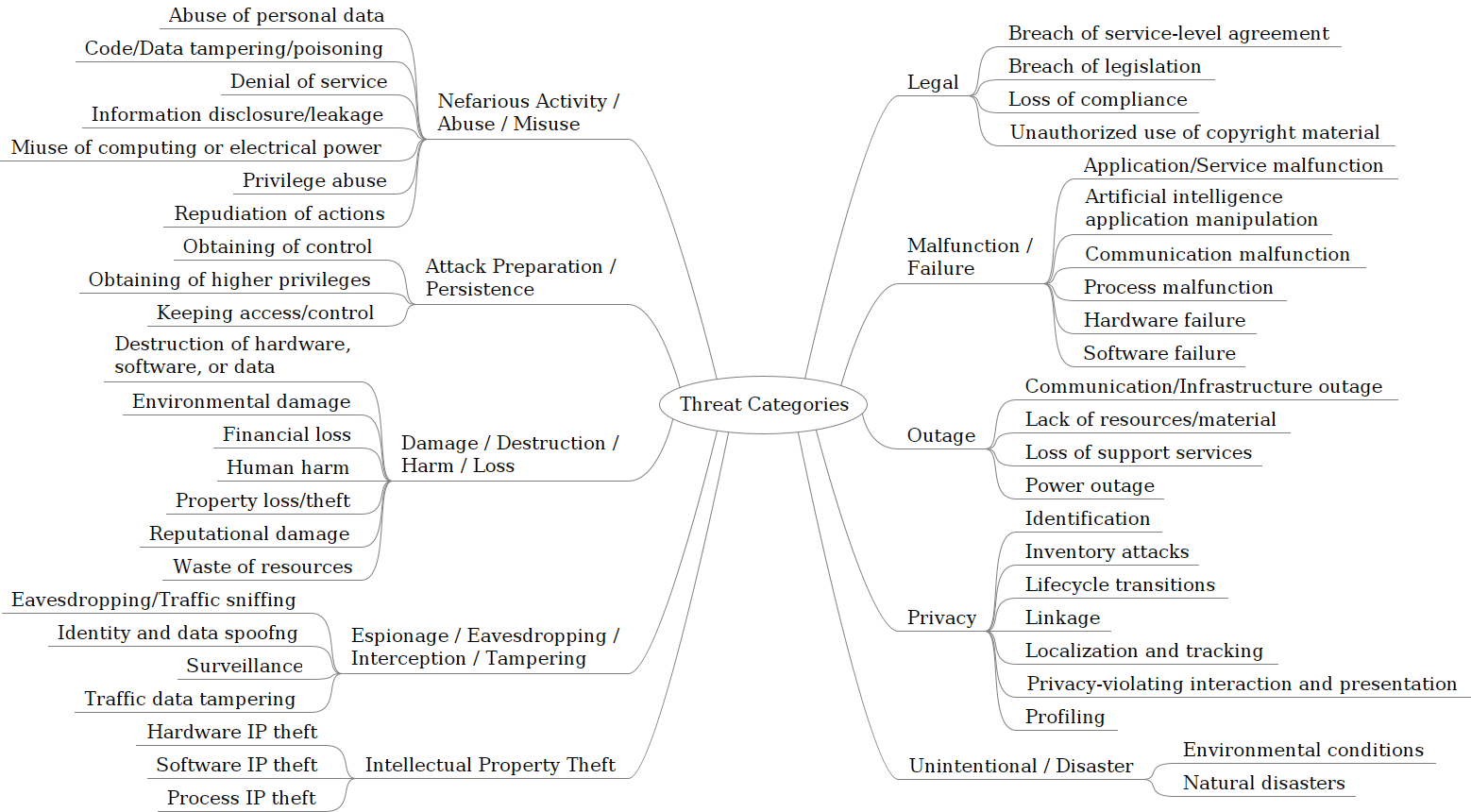}	
	}
	\caption{An overview of the grouped threat categories.}
	\label{fig:Threats}
\end{figure*}

In this section, the threats to IIoT devices are analyzed. The ENISA defines a threat as ``any circumstance or event with the potential to adversely impact an asset through unauthorized access, destruction, disclosure, modification of data, and/or denial of service'' \cite{Threat}. In order to better understand and assess threats, we categorized and grouped them according to their impact. The collected threat categories were identified based on the preceding asset analysis and are presented in Figure~\ref{fig:Threats}. The various threat groups are briefly outlined below.

\subsection{Nefarious Activity / Abuse / Misuse}
The first group summarizes rather generic threats from nefarious activity, abuse, and misuse. Abuse of personal data is the use of personal information in a manner for which it was not intended, for example, a company selling ones email address to advertisers. Another threat is tampering with data or also poisoning of data, which can occur in the context of machine learning. Any threat that compromises availability is summarized as denial of service. The impact of these threats is significant in ICSs because they might lead to a production stop and, thus, cause financial loss or even worse in critical infrastructures. Information disclosure and leakage is sharing of data that is intended to be confidential. Misuse of computing or electrical power can be caused by malware or the inappropriate use of the device by employees. Computing power can be misused by botnets that launch Distributed Denial of Service (DDoS) attacks, mine cryptocurrencies, or spread spam. Employees might use the device to charge their phone, affecting the functionality of the device. Privilege abuse is a threat that arises from employees who maliciously use their privileges. About two out of three incident are financially motivated; further reasons are fun, grudge and espionage \cite{DBIR}. The last category of this group is repudiation of actions. In case of an incident, investigators try to reconstruct the exact procedure, for example, by analyzing log files. Attackers could manipulate or delete them in order to remain
undetected. However, many IIoT devices do not uniquely identify users so far, which makes it easier for them to deny it.


\subsection{Attack Preparation / Persistence}
This group summarizes threats in which attackers gain unauthorized access to the device or a resource to prepare further attacks. A common technique is privilege escalation to obtain either initial access or higher privileges. Afterwards, adversaries might try to keep access across restarts, updates, or factory resets. 

\subsection{Damage / Destruction / Harm / Loss}
Damage can arise in many ways. Attackers can destroy hardware, software, or data of the device. For example, ransomware can destroy data by encrypting it, software can be erased by deleting it requiring a reinstallation (e.g., the malware Brickerbot did it this way \cite{BrickerBot}), and hardware can be destroyed by actuator malfunction, a short circuit, or vandalism. Damage can also exceed device borders. Humans can be injured by attacks compromising safety such as the aforementioned example actuator malfunction and by consequential damage from attacks on the power grid. Additionally, environmental damage can result, as demonstrated by an attack with simple radio signals on Maroochy Water Services that discharged 800,000 liters of sewage to local parks and rivers \cite{Maroochy}. Furthermore, financial loss can occur to manufacturers through product piracy, to owners through production downtime, and to users through theft of credit card information. Manufacturers and operators also fear damage to their company's image. Many manufacturers do not disclose vulnerabilities, operators publish incidents sketchily, and it is also suspected that only a fraction is made public.

\subsection{Espionage / Eavesdropping / Interception / Tampering}
In ICSs, encrypted communication is still rather rare. Industrial espionage by eavesdropping on communications is thus simpler. It also facilitates tampering with data. Payload data such as sensor values and commands can be manipulated, and identities can be spoofed. For instance, attackers could masquerade as the device and send false data
to PLCs or cloud services. In addition, the variety of sensors such as microphones or cameras enables surveillance, for example, to monitor the behavior and activities of employees. 

\subsection{Intellectual Property Theft}
The significance of IP has already been described in Section~\ref{sec:Assets}. The primary source of threats to hardware, software, and process IP is from competitors.

\subsection{Legal}
This group considers legal implications including breach of service-level agreements, breach of legislation, and loss of compliance. An increasing number of laws have recently been drafted for connected devices. For example, a law in California requires to have unique preprogrammed passwords for each device \cite{Cali}. Germany also passed a law requiring manufacturers of digital devices to provide updates \cite{UpdatePflicht}. Consequently, manufacturers must continuously check whether the legal situation for their devices has changed. Last, manufacturers use a variety of third-party components. They must be careful not to use protected material without authorization. 

\subsection{Malfunction / Failure}
One of the most famous attacks that caused application manipulation is Stuxnet, in which the speed of centrifuges was changed, while hiding it from monitoring systems \cite{Stuxnet}. Two recent examples are TRITON \cite{Triton} and Industroyer \cite{Industroyer} that were specifically created for OT devices and protocols. AI methods are currently used in cloud services, edge devices and even smart sensors. Special attention should be paid to them as they can increase the number of threats. Attacks could compromise and limit AI results, reduce their effectiveness, and lead to misclassification by providing adversarial examples. In addition to the application, the communication can also be manipulated. In this case, it is not about single bytes, but rather about entire packets. Gateways, for example, are the link between sensors and cloud services. Occasionally, packets may be redirect or not forwarded at all. Further threats may arise from hardware and software failures; components can fail due to age or quality issues and software may contain bugs.

\subsection{Outage}
This group summarizes outages that affect large parts of an ICS. The power supply is a basic requirement; an abrupt interruption could lead to serious consequences. An outage in communication between the numerous devices could have similar outcomes. However, it is sufficient if a single production support system, such as a logistics service, fails. Furthermore, downtime can occur due to lack of materials. 

\subsection{Privacy}
Privacy in the IoT remains a hot research topic. Due to the complexity of the topic, this section follows the highly regarded paper by Ziegeldorf et al. about privacy threats in the IoT  \cite{PrivacyThreats}. The threats, that mainly concern users, are briefly summarized below. First, the threat identification arises when an identifier can be linked with an individual and data about him/her. This includes the identification of humans as well as devices, for example, by fingerprinting. Second, as devices become more interconnected, they can be queried over the network, allowing attackers to gather information and characteristics about existing devices, called inventory attacks. The resulting inventory list of a factory could be interesting for competitors, for example. Another threat occurs during life cycle transitions. When device users change, the (sensitive) data of the previous user is often still there; a function for disposal is also often missing for full data wipe. The possibility to link two or more previously separated systems poses a further threat. The linkage of their data sources may reveal (truthful or erroneous) information without consent of the user. There are several ways to locate and track users through the device or an associated service. This may be necessary for the functionality of the application, but it also introduces threats; for example, the performance of employees can be tracked. The way users interact with IoT devices and the information they present in response can also result in the disclosure of sensitive data. For example, it might not be possible to privately interact with a voice-controlled device in public space. Last, there is a threat of user profiling to correlate their interests or behavior with other profiles and data.

\subsection{Unintentional / Disaster}
Unfavorable conditions can affect the functionality of devices. It matters little whether these are caused by environmental factors or disasters, if only the impact is considered. Threats can arise from the operation of a device outside of the specified parameters, for example, in unapproved temperature range (e.g., due to fire or lack of switching cabinet cooling) or voltage range (e.g., due to lightning strike or fluctuations in the power supply). Other categories are mechanical stress (earthquake or misuse by employees), pollution (dust), and corrosion.

\section{Attack and Weakness Categorization}
\label{sec:Attacks}

\begin{figure}[ht!]
	\centering
	\resizebox{0.5\textwidth}{!}{
	\begin{tikzpicture}[scale=1., every node/.style={scale=1.0}, mynode/.style={fill=#1,minimum height=0.5cm,minimum width=2cm,rounded corners, draw=gray,very thick,font=\color{white} \small}, mynode/.default=gray]
		\newcommand{\cloud}[3]{%
			\draw[thick, fill=#3, scale=0.5] (#1-1.6,#2-0.7) .. controls (#1-2.3,#2-1.1)
			and (#1-2.7,#2+0.3) .. (#1-1.7,#2+0.3) .. controls (#1-1.6,#2+0.7)
			and (#1-1.2,#2+0.9) .. (#1-0.8,#2+0.7) .. controls (#1-0.5,#2+1.5)
			and (#1+0.6,#2+1.3) .. (#1+0.7,#2+0.5) .. controls (#1+1.5,#2+0.4)
			and (#1+1.2,#2-1) .. (#1+0.4,#2-0.6) .. controls (#1+0.2,#2-1)
			and (#1-0.2,#2-1) .. (#1-0.5,#2-0.7) .. controls (#1-0.9,#2-1)
			and (#1-1.3,#2-1) .. cycle;
		}
					
		{
			\def \rad {3.68}
			\draw[-{Triangle Cap[cap angle=60,bend]},line width=.3cm,gray!70] (0,\rad) arc (90:-213.5:\rad);
			
			\node[mynode=gray!70] at (90-0/7:\rad) {Design};
			\node[mynode=gray!70] at (90-360/7:\rad) {Production};
			\node[mynode=gray!70] at (90-720/7:\rad) {Distribution};
			\node[mynode=gray!70] at (90-1080/7:\rad) {Setup};
			\node[mynode=gray!70] at (90-1440/7:\rad) {Operation};
			\node[mynode=gray!70] at (90-1800/7:\rad) {Maintenance};
			\node[mynode=gray!70] at (90-2160/7:\rad) {End of Life};
			
			\node at (-2,1.55) {\includegraphics[width=.8cm]{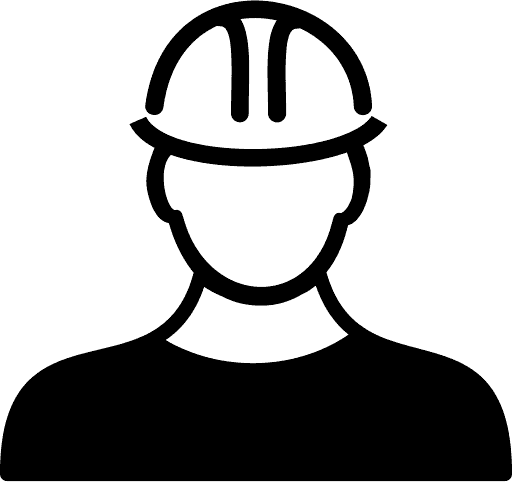}};
			\node[text width=1cm, text centered, scale=0.75] at (-2, 1.05) {User};
			
			\cloud{2}{4.5}{black!5}
			\node[text width=2cm, text centered] at (0.8, 2.4) {\footnotesize Cloud};
			\node[text width=2cm, text centered] at (0.8, 2.15) {\footnotesize Services};
			
			\node at (-0.8,2.5) {\includegraphics[width=.8cm]{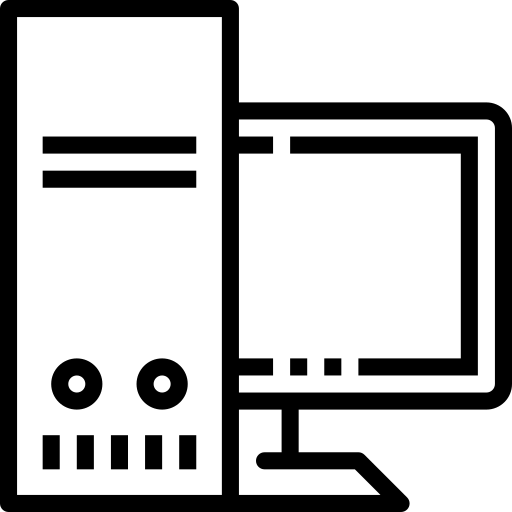}};
			\node[text width=1cm, text centered, scale=0.75] at (-0.8, 1.95) {SCADA};
			
			\node[inner sep=0pt] at (2,1.55) {\includegraphics[width=.8cm]{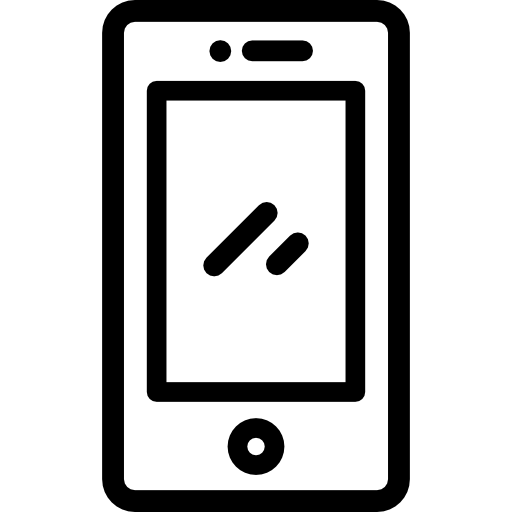}};
			\node[text width=1cm, text centered, scale=0.75] at (2, 1.0) {Phone};
			
			\draw[ultra thick,gray] (-1.7,3.2) -- (-1.2, 3.7);
			\draw[ultra thick,gray,fill=white] (-1.2, 3.7) circle (1mm);
			\draw[ultra thick,gray] (-1.7,3.2) -- (0.45, 2.88);
			\draw[ultra thick,gray,fill=white] (0.45, 2.88) circle (1mm);
			\draw[ultra thick,gray] (-1.7,3.2) -- (-0.82, 0.73);
			\draw[ultra thick,gray,fill=white] (-0.82, 0.73) circle (1mm);
			\draw[ultra thick,gray] (-1.7,3.2) -- (-1.6, 1.63);
			\draw[ultra thick,gray,fill=white] (-1.6, 1.63) circle (1mm);
			\draw[ultra thick,gray] (-1.7,3.2) -- (-2.75, 2.75);
			\draw[ultra thick,gray,fill=white] (-2.75, 2.75) circle (1mm);
			\draw[ultra thick,gray,fill=white] (-1.7,3.2) circle (0.3cm);
			
			\def \arrX {0.9}
			\draw[] (-\arrX,0.03) edge[MyArrow=double] (-\arrX-0.7,0.96);
			\draw[] (0.25,0.26) edge[MyArrow=double] (0.5,1.75);
			\draw[] (-0.25,0.26) edge[MyArrow=double] (-0.5,1.75);
			\draw[] (\arrX,0.03) edge[MyArrow=double] (\arrX+0.7,0.96);
		}
		
		{
			\def \devX {-1.15}
			\def \devY {-2.6}
			\def \coff {0.5}
			\draw[thick] (0,\devY-\devX) circle (-\devX+0.6);
			\draw[thick, rounded corners] (\devX,\devY) rectangle +({-2*\devX},{-2*\devX});
			\draw[thick] (\devX,{\devY-4/3*\devX}) -- +({-2*\devX},0);
			\draw[thick] (\devX,{\devY-2/3*\devX}) -- +({-2*\devX},0);
			\draw[] ({\devX-2/3*\devX+0.1},{\devY-2*\devX}) -- +(0,{2/3*\devX});
			\draw[] ({\devX-4/3*\devX-0.12},{\devY-2*\devX}) -- +(0,{2/3*\devX});
			\def \pin {0.28}
			\draw[thick] ({\devX-1/5*\devX},{\devY-2*\devX}) -- +(0,{\pin});
			\draw[thick] ({\devX-3/5*\devX},{\devY-2*\devX}) -- +(0,{\pin});
			\draw[thick] ({\devX-5/5*\devX},{\devY-2*\devX}) -- +(0,{\pin});
			\draw[thick] ({\devX-7/5*\devX},{\devY-2*\devX}) -- +(0,{\pin});
			\draw[thick] ({\devX-9/5*\devX},{\devY-2*\devX}) -- +(0,{\pin});
			
			\draw[thick] ({\devX-1/5*\devX},{\devY-\pin}) -- +(0,{\pin});
			\draw[thick] ({\devX-3/5*\devX},{\devY-\pin}) -- +(0,{\pin});
			\draw[thick] ({\devX-5/5*\devX},{\devY-\pin}) -- +(0,{\pin});
			\draw[thick] ({\devX-7/5*\devX},{\devY-\pin}) -- +(0,{\pin});
			\draw[thick] ({\devX-9/5*\devX},{\devY-\pin}) -- +(0,{\pin});
			
			\draw[thick] ({\devX-\pin},{\devY-1/5*\devX}) -- +({\pin},0);
			\draw[thick] ({\devX-\pin},{\devY-3/5*\devX}) -- +({\pin},0);
			\draw[thick] ({\devX-\pin},{\devY-5/5*\devX}) -- +({\pin},0);
			\draw[thick] ({\devX-\pin},{\devY-7/5*\devX}) -- +({\pin},0);
			\draw[thick] ({\devX-\pin},{\devY-9/5*\devX}) -- +({\pin},0);
			
			\draw[thick] ({-\devX},{\devY-1/5*\devX}) -- +({\pin},0);
			\draw[thick] ({-\devX},{\devY-3/5*\devX}) -- +({\pin},0);
			\draw[thick] ({-\devX},{\devY-5/5*\devX}) -- +({\pin},0);
			\draw[thick] ({-\devX},{\devY-7/5*\devX}) -- +({\pin},0);
			\draw[thick] ({-\devX},{\devY-9/5*\devX}) -- +({\pin},0);
			\node[text width=3cm, align=center] at (0,\devY-1/3*\devX)  {\small Hardware};
			\node[text width=3cm, align=center] at (0,\devY-3/3*\devX)  {\small Firmware};
			\node[text width=3cm, align=center] at (0,\devY-5/3*\devX)  {\small App$1$~~...~~App$N$};
		}
		
		{
			\newcommand{\attack}[5]{%
				\filldraw[fill=blue1!45, draw=blue1!80, thick, rounded corners] (#1,#2) rectangle +(#3,0.6);
				\node[text width=#3cm, align=center] at ({#1+0.5*#3},#2+2/3*0.5+0.08)  {\footnotesize #4};
				\node[text width=#3cm, align=center] at ({#1+0.5*#3},#2+1/3*0.5)  {\footnotesize #5};
			}

			\attack{-3.3}{-2.95}{2}{Chip, PCB, and}{Device Attacks}
			\node[rotate=90] at (-1.5,-2.2) {\includegraphics[width=.7cm]{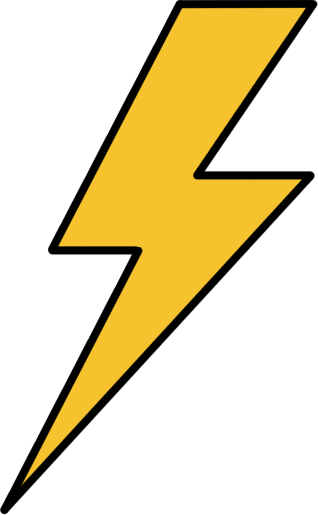}};
			
			\attack{-3}{-1.75}{1.55}{Firmware}{Attacks}
			\node[rotate=90] at (-1.5,-1.1) {\includegraphics[width=.7cm]{lightning.png}};
			
			\attack{1.65}{-1.8}{1.75}{Cryptography}{Attacks}
			\node[rotate=-45] at (1.35,-1.35) {\includegraphics[width=.7cm]{lightning.png}};
			
			\attack{1.4}{-0.1}{1.75}{Application}{Attacks}
			\node[rotate=0] at (1.25,0.1) {\includegraphics[width=.7cm]{lightning.png}};
			
			\attack{0.2}{0.85}{1.35}{Network}{Attacks}
			\node[rotate=0] at (0.1,0.95) {\includegraphics[width=.7cm]{lightning.png}};
			
			\attack{-3.95}{0.85}{1.35}{User}{Behavior}
			\node[rotate=90] at (-2.7,1.5) {\includegraphics[width=.7cm]{lightning.png}};
			
			\attack{-3.7}{0.05}{1.45}{Credential}{Attacks}
			\node[rotate=135] at (-1.85,0.35) {\includegraphics[width=.7cm]{lightning.png}};
			
			\attack{2.75}{3.5}{1.75}{Supply Chain}{Attacks}
			\node[rotate=0] at (2.5,3.6) {\includegraphics[width=.7cm]{lightning.png}};
			
			\attack{-4.2}{3.}{1.75}{Ecosystem}{Weaknesses}
			\node[rotate=90] at (-2.25,3.55) {\includegraphics[width=.7cm]{lightning.png}};
		}
		
	\end{tikzpicture}	
	}
	\caption{IoT devices can be attacked during the entire product life cycle. The various types of attacks on an IoT device, illustrated in the circle, are shown in the blue boxes.}
	\label{fig:Attacks}
\end{figure}

There are countless ways to attack an IIoT device and new attack techniques are also constantly being discovered. Therefore, we categorized common attack techniques and weaknesses (see Figure~\ref{fig:Attacks}). The previously conducted decomposition of a device into its individual components in Section~\ref{sec:Components} was leveraged here to get a comprehensive understanding of the attack surface. Below, these categories are described by exemplary attack vectors throughout the life cycle of an IIoT device, namely design, production, distribution, setup, operation, maintenance, and end of life. Note that not all attacks can be clearly assigned to a single category.

\subsection{Hardware Attacks}

A closer examination of hardware attacks revealed that they can be subdivided into chip-, PCB- and device-level attacks. It should be noted here that this category is different from the commonly used designation \textit{physical attacks}, as physical access is not always necessary.

\paragraph{Chip-Level Attacks}
Attacks against chips, such as microcontrollers and custom ASICs, start at the design stage and also during production. Various actors have the ability to maliciously modify the design and insert a hardware Trojan or backdoor, for example. This can be done by an untrusted IP vendor, foundry, or design facility. Another threat source may be a compiler or Computer-Aided Design (CAD) tool. For example, a modified version of Apple's Xcode infected thousands of iOS apps with malware \cite{Xcode}. The facilities involved also have the ability of IP theft, which enable product piracy. There is still the possibility of reverse engineering in the foundry or later in operation, if someone has no access to the design files. There are further attack opportunities from distribution to the end of life. First, it is possible to tamper with data in chips by exploiting physical access. Such an attack was demonstrated at the Chaos Communication Congress in late 2019, when attackers were able to reroute orders from a microcontroller distributor. After the installation of a backdoor, the controllers could be shipped to the actual customers without them noticing \cite{Roth}. Furthermore, people with (temporary) physical access to the chip can conduct microprobing or fault injection attacks. The former uses microscopic needles to probe internal wires, the latter deliberately injects a fault in a system to change its behavior. Both use invasive attack methods, although fault injection is also possible with semi-invasive (e.g., focused laser beam) and non-invasive (e.g., clock and voltage glitching) attacks \cite{HWAttacks}. These techniques are used to gain secret information or bypass security features, for instance. Another possibility to extract secrets are side-channel attacks. They observe parametric behaviors, such as power consumption or timing information, of a specific implementation of an algorithm to leak encryption keys, for example.

\paragraph{PCB-Level Attacks}
First, attackers might be interested in the PCB design to understand or clone a product causing a threat to IP. They can get the design by stealing the CAD files or by utilizing reverse engineering techniques. Second, a backdoor can be implanted by inserting a malicious component on the PCB or by replacing an existing one. There are several ways to accomplish the former \cite{PCBhacking}. Attackers must initially insert an extra component into one of the different design files. In the second step, it must be mounted on the PCB. This can take place, for example, during production, at repair and rework stations, or after the PCB has been delivered to a warehouse. Another option is to add a completely new component. The project TPM Genie demonstrated that TPMs can be attacked by an interposed device \cite{TPMGenie}. PCBs usually contain interfaces for verifying the design and testing. The commonly integrated protocol JTAG allows to program memory or debug controllers, among others. Attackers can exploit such interfaces to dump the firmware, resulting in an IP theft, or overwrite local storage, enabling firmware and data manipulation.

\paragraph{Device-Level Attacks}
The third group contains more general attack vectors and weaknesses. It includes sabotage attacks that can be any physical change to hardware that has a malicious impact. Threats can also arise unintentionally. A service technician might replace a burnt out PCB or a defective engine with a spare part that was not purchased from the original manufacturer for price reasons resulting in faulty operation, for example. There are other attack possibilities depending on the specific component, e.g., some are vulnerable to magnetic field attacks. One such component are displays. A common weakness occurring in IIoT devices is permissive displayed information to everyone (e.g., the firmware version), if authentication is required at all. Another component are USB ports, which allow insertion of malicious USB sticks that either imitate a keyboard to inject commands or destroy the port or entire device by putting a high voltage on the lines.

\subsection{Firmware Attacks}

The firmware enables several ways to attack a device. Attackers can exploit vulnerabilities in one of its components, for example, a network stack. Within a short time, dozens of vulnerabilities were discovered in famous IoT and OT TCP/IP stacks leading to DoS or Remote Code Execution (RCE); the findings were named as INFRA:HALT, NAME:WRECK, NUMBER:JACK, AMNESIA:33, and Ripple20  \cite{InfraHalt,NameWreck,NumberJack,Amnesia33,Ripple20}. If a vulnerable firmware is already fixed, attackers may try to downgrade it to a previous version with security flaws. Another possibility is to utilize the firmware update mechanism to install a manipulated version including a backdoor, for example. Especially the hardware infrastructure including the boot process is often vulnerable, as concluded by \cite{Thermostat}.

\subsection{Application Attacks}

There are numerous ways to attack applications. However, the techniques depend on the utilized technologies such as web servers, databases, and used programming language. A few of these techniques are described below. A (remote) code execution attack exploits vulnerabilities in a running process, allowing the execution of any instruction on a system. Recently, dozens of vulnerabilities were found in IoT and OT RTOSs exploiting bad memory allocations \cite{BadAlloc}. Code injection attacks that take advantage of insufficiently sanitized input have a similar goal, e.g., SQL injection. Web applications running on the IIoT devices are also exposed to typical vulnerabilities such as Cross-Site Scripting (XSS), broken access control, and insecure deserialization \cite{OWASP}. Furthermore, for CPSs, data injection attacks should be considered, e.g., from spoofed sensor values. There are also attacks against AI applications. It is conceivable to attack machine learning models and data sets during their design or operation. Models can be sabotaged or poisoned and data sets or their labels can be manipulated in order to reduce accuracy, for example.

\subsection{Cryptography Attacks}

Cryptography is a rather cross-disciplinary domain, as it is used for encrypted local storage, authentication, and secure communication, among others. In order to abstract from the various applications, attacks against cryptography have been grouped into this category. A common weakness is the use of deprecated cryptographic algorithms. Especially OT devices with a long lifetime employ these obsolete algorithms, such as the Data Encryption Standard (DES). This can also be suspected from the fact that still many modern microcontrollers implement hardware accelerators for those algorithms. In addition, algorithm-specific parameters, such as the key size, are often selected incorrectly. Examples include a poorly chosen curve in elliptic curve cryptography, and a weak block cipher mode for symmetric algorithms. Cryptography depends heavily on random numbers, which potentially introduces further weaknesses. These include weak pseudo-random number generators, lack of sufficient entropy, and the reuse of nonces (i.e., number that can be used only once). When handling passwords, design flaws can also occur that facilitate brute force or rainbow table attacks. 

\subsection{Network Attacks}

The integration of network interfaces into OT devices has significantly increased their attack surface. One of the major concerns of ICS operators are (D)DoS attacks, as unavailability may cause production downtime, which in turn results in financial damage. In addition to the classic attack techniques, attacks on wireless communication technologies should also be considered, e.g., radio jamming. IIoT devices can be the target of attacks themselves as well as being used to launch attacks against SCADA systems or cloud services. Another risk are the various automation protocols used within OT systems that were designed decades ago and do not support security controls. As a result, man-in-the-middle and replay attacks can often be conducted because of missing authentication and unencrypted communication. Last but not least, network interfaces enable attack preparation attacks, such as port scans or device fingerprinting. 

\subsection{User Behavior}

Users can introduce threats at different stages in the device life cycle. During setup, technicians might misconfigure the device, for example, by disabling security features, or neglect to create a configuration backup and keep this secure. Afterwards, operators might install malware, for instance, to mine cryptocurrencies. The mishandling of warnings and errors, as well as the incorrect use of the device in general, can also lead to further threats. Likewise, errors can occur during maintenance. Technicians are usually responsible for installing firmware updates, as automatic over-the-air updates are problematic due to safety requirements. They are also responsible for deleting all data at the end of life.

\paragraph{Credential Attacks}
Credentials are an important asset of users and, therefore, a major target of attackers. With some exceptions, such as improper storage and cleartext transmission, users are often the weak point to get them. Contrary to recommendations, default and simple passwords are still used, as well as shared passwords with colleagues. Many users also disregard social engineering attacks such as phishing or shoulder surfing.

\subsection{Ecosystem Weaknesses}

This category summarizes weaknesses in the ecosystem that arise due to incomplete system design and the heterogeneous structure of the IIoT. Some manufacturers still do not provide software updates, and if they do, they are rare. Especially IIoT device manufacturers rarely implement vulnerability disclosure policies for fear of damaging their reputation. Further bad practices include hard-coded passwords, developer backdoors, and compromisable procedures for lost credentials. Another vulnerability is the implicit trust between components within the device and also in the entire ecosystem. The latter are still far from implementing the goal of zero trust. A huge challenge is interoperability; solutions are complicated by the numerous legacy standards in ICSs. Other devices and (third-party) IoT services for device life cycle management, telemetry, and analytics are therefore blindly trusted. 

\subsection{Supply Chain Attacks}

Several supply chain attacks have already been mentioned in the previous categories. For the sake of clarity, these are again summarized and expanded in this section. A frequent target is malicious modification of functionality by hardware, software or data. Hardware logic insertion occurs during design and production stage, and replacing a hardware component with a malicious one is feasible during production and in all later stages. The manipulation or insertion of software code is conceivable in all life phases, e.g., by malicious third-party libraries or by a manipulated firmware. Data used to train machine learning models, for example, can also be compromised. IP theft of hardware and software is also possible in all stages; in the case of hardware, this is particularly achievable during design and production, as these steps are often outsourced to external contractors. Outsourcing the production additionally enables cloning of provisioning data and unauthorized overproduction; discarded or defective equipment could moreover end up on the gray market. A further threat to production is the use of counterfeit components or generally inferior material.

\section{Recommended Procedure for Analyzing Threats}
\label{sec:Procedure}

Last, we recommend a procedure for the threat analysis of IIoT devices. For this purpose, the previously discussed aspects are revisited and put in order. Note, this recommendation can be seen as complementary to existing Threat Analysis and Risk Assessment (TARA) methodologies, such as IEC 62443-3-2, in order to facilitate a device-oriented procedure.

\renewcommand\thesubsectiondis{\arabic{subsection}.}
\subsection{System Analysis}

The first step is to analyze the IIoT device in depth. All hardware and software components and data files, as described in Section~\ref{sec:Components}, should be collected. This enables a component-by-component threat and vulnerability analysis later on. In addition, requirements and (environmental) conditions should be gathered. Requirements may originate, for example, from the operating location, industrial sector, or use in critical infrastructures.

\subsection{System Interaction Overview}

As previously stated, connectivity is a central element of IIoT devices. Therefore, an overview of all interactions with the device is necessary for a thorough threat analysis. This diagram should contain all actors including humans, cloud services, and other machines. This can be supported by an additional use case diagram to capture when and why an actor interacts with the system. 

So far, we have only referred to users in general terms; especially in the case of industrial systems, it is sensible to differentiate between them according to their role. For example, the commissioning, operation, and maintenance personnel are often not the same. The goal is to grant each group only the least required privileges. The authorization of certain network interfaces should also be considered. This is especially important for industrial protocols, such as PROFINET. While most IoT applications allow the implementation of security measures manually, it is not possible with these proprietary protocols, as compatibility with other manufacturers must be maintained.

\subsection{Asset Identification}

The third step is the identification of assets based on the various types presented in Section~\ref{sec:Assets}. The assets can be best identified by considering the interests of the different stakeholders; again, the different roles of users should be distinguished. This step helps to understand what values to each party. It also allows to rank the criticality of assets, which can be used for defining the impact of attacks.

Furthermore, it is useful to determine the security goals in general and also per asset. This can support the development of countermeasures later. For CPSs, availability is often most important for safety reasons. For edge devices, confidentiality could be rated higher, as they aggregate information. However, in low-power IIoT devices that must comply with real-time requirements, it is not always feasible to implement the most secure countermeasure. Thus, this initial assessment can be used to choose the right measure. For example, sensitive data in transit (e.g., credentials) will be encrypted, less critical data (e.g., sensor values) will only be protected against manipulation using a message authentication code.

\subsection{Threat Source Identification}

Threat sources can be best identified from the perspective of the attacker to find out what they are targeting and why. There are several types of attackers with different capabilities, attack techniques and motives. We classified various types of threat sources and their respective intentions in \cite{MyTA}. This is useful for deliberately including or excluding types of attacks. For IIoT devices in critical infrastructures, the more complex hardware and supply chain attacks should be addressed.

\subsection{Threat and Vulnerability Identification}

\begin{figure}[ht!]
	\centering
	\resizebox{0.5\textwidth}{!}{
		\begin{tikzpicture}
			\newcommand{\leaf}[5]{%
				\draw (#1,#2) rectangle +(#3,0.6);
				\node[text width=#3cm, align=center] at ({#1+0.5*#3},#2+2/3*0.5+0.09)  {\footnotesize #4};
				\node[text width=#3cm, align=center] at ({#1+0.5*#3},#2+1/3*0.5-0.01)  {\footnotesize #5};
			}
			\newcommand{\leafs}[4]{%
				\draw (#1,#2) rectangle +(#3,0.6);
				\node[text width=#3cm, align=center] at ({#1+0.5*#3},#2+0.275) {\footnotesize #4};
			}
			
			\leaf{-1.25}{0}{2.5}{Obtain}{Manufacturer Key}
			\draw (0,0) -- (0,-0.3);
			
			\def \lvlone {-1.2}
			\def \ydiff {-0.92}
			\def \lvltwo {\lvlone+\ydiff}
			\def \lvlthree {\lvlone+2*\ydiff}
			\def \lvlfour {\lvlone+3*\ydiff}
			\def \lvlfive {\lvlone+4*\ydiff}
			\def \lvlsix {\lvlone+5*\ydiff}
			\def \lvlseven {\lvlone+6*\ydiff}
			\def \xstart {-5}
			\def \xdiff {2.7}
			\def \xoff {0.5}
			\def \wid {2}

			\draw (\xstart+1,-0.3) -- (\xstart+3*\xdiff+1,-0.3);
			\draw (\xstart+1,-0.3) -- (\xstart+1,-0.6);
			\draw (\xstart+1+\xdiff,-0.3) -- (\xstart+1+\xdiff,-0.6);
			\draw (\xstart+1+2*\xdiff,-0.3) -- (\xstart+1+2*\xdiff,-0.6);
			\draw (\xstart+1+3*\xdiff,-0.3) -- (\xstart+1+3*\xdiff,-0.6);
			
			\leaf{\xstart}{\lvlone}{\wid}{Dump}{Memory}
			\leaf{\xstart+\xdiff}{\lvlone}{\wid}{Attack Key}{Processing}
			\leaf{\xstart+2*\xdiff}{\lvlone}{\wid}{Attack}{Supply Chain}
			\leafs{\xstart+3*\xdiff}{\lvlone}{\wid}{...}
			
			\leaf{\xstart+\xoff}{\lvltwo}{\wid}{Dump Memory}{using RCE}
			\leaf{\xstart+\xoff}{\lvlthree}{\wid}{Exploit Physical}{Access}
			\draw (\xstart+0.25,\lvlone) -- (\xstart+0.25,\lvlthree+0.3);
			\draw (\xstart+0.25,\lvltwo+0.3) -- (\xstart+\xoff,\lvltwo+0.3);
			\draw (\xstart+0.25,\lvlthree+0.3) -- (\xstart+\xoff,\lvlthree+0.3);
			
			\leaf{\xstart+2*\xoff}{\lvlfour}{\wid}{Access External}{Memory}
			\leaf{\xstart+2*\xoff}{\lvlfive}{\wid}{Access Internal}{Memory}
			\draw (\xstart+0.25+\xoff,\lvlthree) -- (\xstart+0.25+\xoff,\lvlfive+0.3);
			\draw (\xstart+0.25+\xoff,\lvlfour+0.3) -- (\xstart+2*\xoff,\lvlfour+0.3);
			\draw (\xstart+0.25+\xoff,\lvlfive+0.3) -- (\xstart+2*\xoff,\lvlfive+0.3);
			
			\leaf{\xstart+3*\xoff}{\lvlsix}{\wid}{Break Code}{Protection}
			\draw (\xstart+0.25+2*\xoff,\lvlfive) -- (\xstart+0.25+2*\xoff,\lvlsix+0.3);
			\draw (\xstart+0.25+2*\xoff,\lvlsix+0.3) -- (\xstart+3*\xoff,\lvlsix+0.3);
			
			\leaf{\xstart+\xdiff+\xoff}{\lvltwo}{\wid}{Side-Channel}{Attack}
			\draw (\xstart+0.25+\xdiff,\lvlone) -- (\xstart+0.25+\xdiff,\lvltwo+0.3);
			\draw (\xstart+0.25+\xdiff,\lvltwo+0.3) -- (\xstart+\xoff+\xdiff,\lvltwo+0.3);
			
			\leafs{\xstart+\xdiff+2*\xoff}{\lvlthree}{\wid}{...}
			\draw (\xstart+0.25+\xdiff+\xoff,\lvltwo) -- (\xstart+0.25+\xdiff+\xoff,\lvlthree+0.3);
			\draw (\xstart+0.25+\xdiff+\xoff,\lvlthree+0.3) -- (\xstart+\xdiff+2*\xoff,\lvlthree+0.3);
			
			\leafs{\xstart+2*\xdiff+\xoff}{\lvltwo}{\wid}{...}
			\draw (\xstart+0.25+2*\xdiff,\lvlone) -- (\xstart+0.25+2*\xdiff,\lvltwo+0.3);
			\draw (\xstart+0.25+2*\xdiff,\lvltwo+0.3) -- (\xstart+\xoff+2*\xdiff,\lvltwo+0.3);
		\end{tikzpicture}	
	}
	\caption{Fraction of an example attack tree on an asset.}
	\label{fig:Tree}
\end{figure}
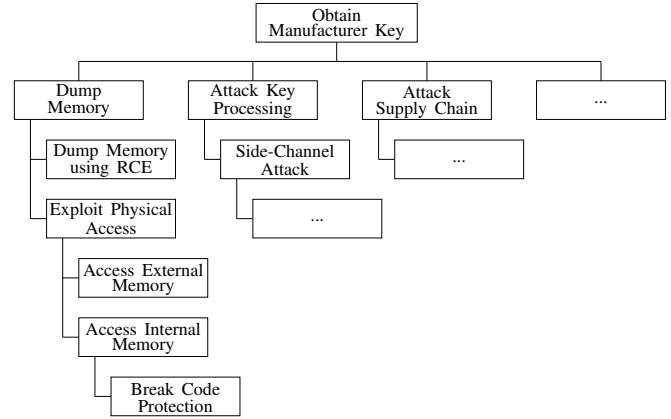

The primary task of a threat analysis is the identification of threats and vulnerabilities. However, the preparatory work from the previous four steps should significantly accelerate and enhance this process. As said before, the system analysis should allow the detection of threats to single hardware components, such as PCBs and actuators. The list of software components can also be used to search for publicly known vulnerabilities. Numerous ways to attack them were presented in Section~\ref{sec:Attacks}. Additionally, it is possible to reflect on how the various threats, presented in Section~\ref{sec:Threats}, may arise. Using attack trees, it is also possible to graphically show how assets can be attacked. Figure~\ref{fig:Tree} shows a sample tree for attacking the manufacturer key that is used for firmware updates, for instance. One option is to dump the memory, which can be achieved either by (remote) code execution or by exploiting physical access. In the latter case, it also depends on where the key is stored; for microcontroller internal memory, a read-out protection may have to be broken.

\captionsetup{font={footnotesize,sc},justification=centering,labelsep=period}%
\begin{table*}[]
	\centering
	\caption{Excerpt of an Example List of Attack Scenarios.}
	\begin{tabular}{lp{3.cm}p{3.4cm}p{1.7cm}ll}
		\toprule
		\multirow{2}{*}{No.} & \multirow{2}{*}{Vulnerability}                                                                   & \multirow{2}{*}{\begin{tabular}[c]{@{}l@{}}Threat\\ (Category)\end{tabular}} & \multicolumn{2}{l}{Attack Vector} & \multirow{2}{*}{Note} \\
		& & & Interface & Action & \\
		\midrule
		1 & \begin{tabular}[c]{@{}l@{}}Default PIN\\ (users did not change it)\end{tabular} & \begin{tabular}[c]{@{}l@{}}Login as administrator\\ (obtaining of control)\end{tabular} & \begin{tabular}[c]{@{}l@{}}Bluetooth app,\\ local HMI\end{tabular} & Authenticate using default PIN & \begin{tabular}[c]{@{}l@{}}PIN can be found\\in the manual\end{tabular} \\
		1.1 &  & \begin{tabular}[c]{@{}l@{}}Moving actuator\\ (application malfunction)\end{tabular} & Bluetooth app & Open/close valve & \\
		1.2 &  & \begin{tabular}[c]{@{}l@{}}Blocking remote control\\ (denial of service)\end{tabular} & Bluetooth app & \begin{tabular}[c]{@{}l@{}}Change control system\\ communication parameter\end{tabular} & \\
		1.3 &  & \begin{tabular}[c]{@{}l@{}}Manipulating user database\\ (data tampering)\end{tabular} & Local HMI & \begin{tabular}[c]{@{}l@{}}View/change/add/delete\\ user accounts\end{tabular} &
		\\
		\bottomrule         
	\end{tabular}
	\label{tab:Scenarios}
\end{table*}
\captionsetup{font={footnotesize,rm},justification=centering,labelsep=period}%

The system interaction and use case diagrams can be utilized to create data flow diagrams in order to identify threats to data in transit. These can be used in combination with threat modeling techniques such as STRIDE \cite{Shostack}, a mnemonic for threats against the security goals authenticity, integrity, non-repudiation, confidentiality, availability, and authorization. Furthermore, penetration testing can be used to discover additional vulnerabilities as well as to verify those already identified and show their severity. 

Documenting the discovered threats is also crucial. We have had the experience that detailed attack scenarios can be better understood in retrospect. Table~\ref{tab:Scenarios} shows an excerpt of a potential attack scenario description. This example lists threats resulting from a default PIN used in an electrical actuator controlling a valve. The PIN is required for authentication at the attached HMI and in a Bluetooth mobile app. An unchanged PIN would allow attackers to authenticate as the administrator. 

\subsection{Vulnerability and Risk Assessment}

The final step is to gather all the information to evaluate the vulnerability and assess the risk. A popular method for vulnerability assessment is the Common Vulnerability Scoring System (CVSS). It incorporates exploitability metrics, such as the attack vector, and the impact on confidentiality, integrity, and availability. The example in Table~\ref{tab:Scenarios} indicates that several threats arise from a single vulnerability. As the impact can vary per interface, an overview of all scenarios is important for a proper scoring. Finally, the risk assessment may include further criteria such as likelihood and (business) impact.

\section{Conclusions}
\label{sec:Conclusions}

Targeted attacks on ICSs, including those in critical infrastructures, increased recently. IIoT devices that merge the previously separated areas of IT and OT additionally increase the attack surface. The consequences of an attack can be dramatic, as OT equipment controls physical processes that, in case of compromised safety, can cause harm to humans, machines, and the environment. Therefore, it is even more essential that IIoT devices are properly secured. However, this is not often the case in reality. One reason is that manufacturers are extensively provided with literature on best practices rather than threat analysis techniques for their devices.

In this paper, we presented a systematic and holistic procedure for analyzing the attack surface and threats of IIoT devices throughout the product life cycle. First, an arbitrary IIoT device was decomposed into its components for this purpose. This itemization of hardware and software components as well as data types is essential to ensure that no attack vectors are overlooked. Afterwards, the assets were analyzed from the perspective of different stakeholders in order to identify everything that is valuable and worth protecting. The provided comprehensive categorization of threats shows an overview of possible threats and their consequences. The attack techniques are almost innumerable and are also constantly expanding. Therefore, we categorized attack techniques and weaknesses that are frequently exploited. This included not only common attack vectors in communications and web applications, but also attacks against the supply chain and the various hardware components. Finally, the threat analysis procedure was described, which enables manufacturers and operators of IIoT devices to identify and evaluate attack vectors. Since the threat categorization considers all assets and the attack categorization addresses all components, the proposed analysis technique seems to be valuable. In the next steps, the procedure will be further validated.

\section*{Acknowledgment}

The research project ``Intelligent Security for Electrical Actuators and Converters in Critical Infrastructures (iSEC)'' is a collaboration of SIPOS Aktorik GmbH, Grass Power Electronics GmbH and OTH Amberg-Weiden. It is supported and funded by the Bavarian Ministry of Economic Affairs, Regional Development and Energy.

\balance

\end{document}